\documentclass[aps,prd,reprint]{revtex4-2}

\usepackage{amsmath,amssymb,physics}
\usepackage{graphicx}
\usepackage{caption}
\usepackage{verbatim}

\usepackage[colorlinks=true,
            linkcolor=blue,
            citecolor=blue,
            urlcolor=blue]{hyperref}

\begin{document}

\title{Heavy-quark spin precession as a magnetic field chronometer}

\author{Dushmanta Sahu}
\email{Dushmanta.Sahu@cern.ch}
\affiliation{Instituto de Ciencias Nucleares, UNAM, Apartado Postal 70-543, Coyoacán, 04510, México City, México}

\author{Captain R. Singh}
\email{captainriturajsingh@gmail.com}
\affiliation{School of Physical Sciences, National Institute of Science Education and Research,
Jatni, Odisha–752050, India}

\begin{abstract}
The lifetime of the strong magnetic field generated in relativistic heavy-ion collisions remains one of the major open questions in the study of the quark--gluon plasma. We propose heavy-quark spin precession as a probe to quantify the magnetic field and its profile in such collisions. We investigate the Larmor evolution of polarized charm and bottom quarks propagating through an expanding quark--gluon plasma in the presence of the magnetic field. We show that realistic magnetic-field strengths and lifetimes generate measurable accumulated precession phases prior to hadronization, leading to characteristic charge-dependent polarization splittings between $\Lambda_c^+$ and $\bar{\Lambda}_c^-$, as well as between $\Lambda_b^0$ and $\bar{\Lambda}_b^0$, through the mixing of transverse and longitudinal polarization components. The predicted signals exhibit distinctive transverse-momentum and quark-mass dependence, providing complementary sensitivity to the magnetic-field lifetime in the charm and bottom sectors. Our results suggest heavy-flavor polarization, especially in the charm sector, as a quantitative probe for the space--time evolution of magnetic fields in relativistic heavy-ion collisions.

\end{abstract}

\maketitle
\section{Introduction}

Relativistic heavy-ion collisions generate the strongest electromagnetic fields that have ever existed in nature, with peak strengths approaching the order of $m_\pi^2$ during the earliest stages of the collision~\cite{Skokov:2009qp, Bzdak:2011yy, Tuchin:2013ie}. While the initial production of these fields is well understood, their subsequent evolution remains one of the major unresolved problems in heavy-ion physics. In the heavy-ion collisions (HICs), fast-moving spectators (protons) induce this strong magnetic field, which rapidly dies out as the nuclei get separated. However, the quark--gluon plasma (QGP), possibly produced in HICs, is proposed to exhibit a finite electrical conductivity and therefore can induce electromagnetic currents that substantially prolong the field lifetime~\cite{Tuchin:2013ie, Stewart:2021mjz}. Therefore, determining the time dependence of the magnetic field is essential for understanding a wide range of electromagnetic phenomena in strongly interacting matter, including anomalous transport, chiral effects, and spin dynamics~\cite{McLerran:2013hla, Kharzeev:2015znc, Gao:2020vbh}.  The concept of spin precession induced by gauge fields has a long history in the investigations on QCD matter. In particular, quark spin precession through the chromomagnetic moment in an effective inhomogeneous color field has been invoked to describe transverse single-spin asymmetries in hadronic collisions~\cite{Abramov:2007sr}. The chromomagnetic interaction of heavy quarks is also a fundamental spin-dependent contribution in Heavy Quark Effective Theory (HQET), appearing at subleading order in the heavy-quark expansion~\cite{Amoros:1997rx, Grozin:2007fh}. These developments motivate the study of spin precession as a probe of gauge-field dynamics. The current study considers the electromagnetic counterpart of this mechanism in relativistic heavy-ion collisions, where the early production and large masses of heavy quarks make their spin evolution particularly sensitive to the space-time evolution of the produced magnetic field.\\

The recent independent observations of the global $\Lambda$ and $\bar{\Lambda}$ polarization at the STAR and ALICE experiments demonstrated the transfer of orbital angular momentum to particle spin through thermal vorticity~\cite{STAR:2017ckg, ALICE:2019onw}. These polarization measurements imply that spin observables can serve as the probe for the rotational dynamics of deconfined QCD matter.  Besides vorticity, however, magnetic fields can also affect the spin dynamics of quark and antiquark, either directly by coupling with the spin~\cite{Guo:2019joy}, or through Larmor precession, causing an initially polarized spin vector to rotate during its propagation through the medium~\cite{Sahu:2026aqw}. Since the accumulated precession angle is proportional to the time integral of the magnetic field, spin observables offer a direct opportunity to probe the space--time evolution of the electromagnetic field rather than merely its initial magnitude.\\

In a previous study, magnetic-field-induced Larmor precession of strange quarks has been proposed as a mechanism for generating charge-dependent polarization splittings between $\Lambda$ and $\bar{\Lambda}$ hyperons~\cite{Sahu:2026aqw}. While this establishes a novel connection between polarization observables and the magnetic field, the strange sector is subject to several theoretical uncertainties. The effective strange-quark mass depends strongly on the properties of the medium, and the observed $\Lambda$ polarization receives contributions from thermal vorticity, shear-induced polarization, resonance feed-down, hadronic re-scattering, and other final-state effects~\cite{Han:2017hdi, Becattini:2016gvu}. These contributions impose a complex phenomenon on the quantitative extraction of the magnetic-field dynamics from strange-hadron measurements. On the other hand, heavy quarks provide a qualitatively different probe of electromagnetic spin dynamics. Charm and bottom quarks are produced predominantly during the initial hard-scattering processes and therefore experience the strongest magnetic fields produced initially just after the collision~\cite{Andronic:2015wma}. Their large masses remain essentially unchanged throughout the evolution of the QGP, thereby reducing theoretical uncertainties associated with medium-dependent thermal masses. As heavy-quark spin-flip processes are suppressed, it leads to comparatively long spin-relaxation times, and causes spin polarization to evolve coherently throughout the deconfined phase via Larmor precession around the magnetic-field direction. Consequently, heavy quarks can retain information about their initial spin orientation over a substantial fraction of the QGP evolution and are particularly well suited for studying coherent magnetic spin precession.\\

An equally important feature of heavy quarks is their complementary sensitivity to the magnetic-field evolution. The accumulated Larmor phase scales can be approximated as;
\[
\Theta \propto \frac{q}{E}\int B(\tau)\,d\tau
\]
which implies an explicit dependence of the spin rotation on the magnetic field profile. Charm quarks, owing to their relatively smaller mass compared to bottom quarks, develop larger precession angles and therefore demonstrate comparatively high sensitivity to the magnetic field. Bottom quarks, although exhibiting smaller spin rotations, remain polarized over comparable timescales and offer an independent probe with a different mass scale. A combined analysis of the charm and bottom sectors, therefore, provides complementary constraints on the lifetime and evolution of the magnetic field that are difficult to obtain from a single quark flavor alone. Moreover, the polarization of heavy-flavor baryons provides an experimentally accessible realization of this approach. In the heavy-quark limit, the spins of the $\Lambda_c$ and $\Lambda_b$ baryons are dominated by the charm and bottom quarks, respectively, implying that a substantial fraction of the heavy-quark polarization is transferred to the final hadron. Moreover, the opposite electric charges of heavy quarks and antiquarks lead to opposite Larmor precession phases in the same magnetic field, generating characteristic charge-dependent polarization splittings between $\Lambda_c^+$ and $\bar{\Lambda}_c^-$, as well as between $\Lambda_b^0$ and $\bar{\Lambda}_b^0$. Such charge-odd polarization patterns are difficult to mimic through conventional vorticity-driven mechanisms and therefore constitute distinctive signatures of electromagnetic spin dynamics.\\

In this work, we investigate the Larmor precession of polarized charm and bottom quarks propagating through an expanding magnetized QGP under conditions relevant to LHC collisions. Employing a Bjorken-expanding medium together with phenomenological models for the magnetic-field evolution, we calculate the accumulated precession phase and its dependence on magnetic-field strength, lifetime, and transverse momentum. We demonstrate that realistic magnetic fields generate measurable spin rotations prior to hadronization, leading to characteristic charge-dependent polarization patterns in heavy-flavor baryons. Our findings suggest heavy-flavor polarization as a quantitative probe for the magnetic field produced in heavy-ion collisions and provide a framework for constraining the lifetime of the field in these ultra-relativistic collisions.

\section{Formalism}

The magnetic field is assumed to be perpendicular to the reaction plane, and along the direction of the net angular momentum (the $y$-axis) of the system. The lifetime of the magnetic field depends sensitively on the electrical conductivity of the QGP and the electromagnetic response of the medium. To investigate the robustness of our approach, we consider three representative magnetic-field evolution scenarios that encompass the range of behaviors broadly discussed in the literature. The chosen magnetic field profiles are as follows;

\begin{equation}
eB(\tau)=
\begin{cases}
eB_{0}
\left[
\dfrac{1+\tau^{2}/\tau_{B}^{2}}
{1+\tau_{0}^{2}/\tau_{B}^{2}}
\right]^{-3/2}
& \text{Vacuum}\\[0.45cm]

eB_{0}\exp\!\left[-(\tau-\tau_{0})/\tau_{B}\right]
& \text{Exponential}\\[0.45cm]

\dfrac{eB_{0}}
{1+\tau^{2}/\tau_{B}^{2}}
& \text{Viscous MHD}
\end{cases}
\label{eq:Bprofiles}
\end{equation}
\noindent
here $eB_{0}$ denotes the initial magnetic-field strength, at the initial time $\tau_0=0.5 ~{\rm fm}$ and $\tau_{B}$ characterizes the timescale for decaying magnetic-field.  Throughout this work, we consider initial magnetic-field strengths in the range $eB_{0}=(1-5)m_{\pi}^{2}$ which encompasses the values typically predicted for Pb--Pb collisions at LHC energies. The characteristic decay time is varied within $\tau_{B}=3-7~\mathrm{fm}$, covering the range from rapidly decaying spectator fields to long-lived fields sustained by the conducting QGP medium.\\

Now, the first case mentioned in Eq.~\ref{eq:Bprofiles} corresponds to the evolution of the magnetic field in the vacuum or in the absence of any kind of conducting medium~\cite{Kharzeev:2007jp,Chen:2021nxs}. Under such a regime, the magnetic field decreases rapidly as the spectator protons recede from the interaction region and therefore represents the shortest possible magnetic-field lifetime. On the other hand, the second profile is a phenomenological exponential parametrization that has been widely employed in studies of electromagnetic effects in heavy-ion collisions~\cite{Voronyuk:2011jd,Deng:2012pc}. The parameter $\tau_{B}$ gets modified in the presence of an evolving conducting medium and effectively controls the lifetime of the magnetic field in a medium. Such a profile allows us to systematically investigate the sensitivity of heavy-quark spin precession across scenarios accessible by varying $\tau_{B}$. The next chosen profile represents the resistive magnetohydrodynamic (MHD) scenario~\cite{Xu:2020sui,Wei:2024lah}, in which the finite electrical conductivity of the QGP induces electric currents that partially sustain the magnetic field. Consequently, the magnetic field lifetime in the medium increases as compared to the exponential decay profile. By comparing predictions from these three magnetic-field evolution scenarios, we assess the sensitivity of heavy-quark spin precession to the electromagnetic history of the medium and identify observables that can discriminate between competing models of magnetic-field evolution.\\

In a system with charged particles, spin dynamics is governed by the Bargmann--Michel--Telegdi (BMT) equation~\cite{bmt}. In the local rest frame of the fluid element, where the electromagnetic field is dominated by its magnetic component and electric field, and kinematics corrections are neglected, the BMT equation reduces to the Larmor precession,

\begin{equation}
\frac{d\mathbf{P}}{d\tau}
=
\boldsymbol{\omega}_L(\tau)
\times
\mathbf{P},
\end{equation}
\noindent
where $\mathbf{P}$ denotes the heavy-quark polarization vector and $\boldsymbol{\omega}_L$ is the instantaneous Larmor frequency. For a heavy quark of electric charge $q_Q$ and energy $E_Q$, the Larmor frequency is defined as;
\begin{equation}
\omega_L^Q(\tau)
=
\frac{g_Q|q_Q|\,eB(\tau)}
{2E_Q}
\end{equation}
here, $g_Q\simeq2$ is the heavy-quark gyromagnetic factor and  $E_Q = \sqrt{p_T^2+p_z^2+m_Q^2}$ is the energy of heavy-quark. The $m_{Q}$ represents the bare mass of the heavy quark, which corresponds to charm $m_c=1.27~{\rm GeV}$ and bottom $m_b=4.18~{\rm GeV}$ quark masses. $q_{Q}$ is the electric charge which is $\frac{2}{3}$e and $\frac{-1}{3}$e for charm and bottom quark respectively.\\

Now, the accumulated Larmor phase acquired by a heavy quark during the QGP evolution is given as;

\begin{equation}
\Theta
=
\int_{\tau_0}^{\tau_f}
\omega_L^Q(\tau)\,
d\tau  
 =
\frac{g_Q|q_Q|\,}
{2E_Q}\int_{\tau_0}^{\tau_f}
eB(\tau)\,
d\tau 
\label{lp1}
\end{equation}
\noindent
where $\tau_f$ is the time when the medium approaches its thermal phase boundary at pseudo-critical temperature  T = 155 MeV, with a 1D Bjorken cooling~\cite{Bjorken1983}. The Larmor precession of charm and bottom quarks (antiquarks) is obtained for the chosen set of magnetic field profiles using Eq.~\ref{lp1}.\\











The physical consequence of Larmor precession is the rotation of the polarization vector in the plane perpendicular to the magnetic field. Taking the magnetic field to be oriented along the $y$ axis, the polarization component parallel to the field remains unchanged, i.e. $P_y^{\,f}=P_y^{\,0}.$, whereas the transverse components evolve as;
\begin{equation}
P_x^{\,f}
=
P_x^{\,0}\cos\Theta
-
P_z^{\,0}\sin\Theta,
\end{equation}
\begin{equation}
P_z^{\,f}
=
P_z^{\,0}\cos\Theta
+
P_x^{\,0}\sin\Theta,
\end{equation}


For heavy antiquarks, the opposite electric charge reverses the precession direction, corresponding to the replacement $\Theta\rightarrow-\Theta$. Consequently,

\begin{align}
P_{x,\bar Q}^{\,f}
&=
P_x^{\,0}\cos\Theta
+
P_z^{\,0}\sin\Theta,\\
P_{z,\bar Q}^{\,f}
&=
P_z^{\,0}\cos\Theta
-
P_x^{\,0}\sin\Theta.
\end{align}

The resulting charge-dependent polarization splittings become

\begin{equation}
\Delta P_x
=
P_{x,\bar Q}^{\,f}
-
P_{x,Q}^{\,f}
=
2P_z^{\,0}\sin\Theta,
\end{equation}

\begin{equation}
\Delta P_z
=
P_{z,Q}^{\,f}
-
P_{z,\bar Q}^{\,f}
=
2P_x^{\,0}\sin\Theta.
\end{equation}

Since the charm and bottom quarks carry electric charge $q_Q$, the same magnetic field that drives spin precession also acts on the quark momentum via the Lorentz force,
\begin{equation}
\frac{dp_x}{d\tau} = \frac{q_Q eB_y}{E}p_z,\qquad
\frac{dp_z}{d\tau} = -\frac{q_Q eB_y}{E}p_x,
\label{eq:momentum}
\end{equation}
where $E$ is the quark energy. Writing $(p_x,p_z)=|\mathbf{p}_\perp|(\cos\phi_p,\sin\phi_p)$, one sees that this force leaves $|\mathbf{p}_\perp|$ exactly unchanged and reduces to a single evolution equation for the momentum azimuthal angle,
\begin{equation}
\frac{d\phi_p}{d\tau} = \frac{|q_Q|eB_y(\tau)}{E}.
\label{eq:phip}
\end{equation}

At $g_s=2$, the BMT equation predicts vanishing anomalous precession, $a_s=(g_s-2)/2=0$. Consequently, if one sets $E=m_s$, the accumulated rotation of the momentum direction becomes exactly equal to the spin precession angle $\Theta$. It suggests that transverse momentum ($p_{\rm T}$) remains unaffected, while the azimuthal angle $\phi_p$ precesses in lockstep with the spin. The baseline polarizations $P_{x}^0(p_{\rm T})$ and $P_{z}^0(p_{\rm T})$ are obtained as the harmonic moments $\langle P_x\sin(2\phi_p)\rangle$ and $\langle P_z\sin(2\phi_p)\rangle$, evaluated at the production angle $\phi_p^0$. As the angle is measured at freeze-out, it is shifted to $\phi_p^0\pm\Theta$, projecting this moment onto the measured angle can be obtained via;
\begin{equation}
\sin[2(\phi_p^0\pm\Theta)]=\sin(2\phi_p^0)\cos(2\Theta)\pm\cos(2\phi_p^0)\sin(2\Theta).
\label{eq:addition}
\end{equation}
Under this assumption, $\sin(2\phi_p)$ is the only non-vanishing azimuthal harmonic; i.e., $\langle\cos(2\phi_p^0)\ rangle=0$. The same assumption is already implicit in reporting only this single moment, and the second term drops out upon averaging, which leaves a single multiplicative dilution factor,
\begin{equation}
D(\Theta) = \cos(2\Theta).
\label{eq:dilution}
\end{equation}

$D(\Theta)$ being an even function of $\Theta$, it suppresses the polarization of particles and antiparticles identically, and therefore it is unable to alter any charge-odd rotation. It is precisely this charge-odd polarization splitting that constitutes the central observable proposed in this present work. Since the splitting is directly proportional to the accumulated precession phase, measuring the polarization differences between $\Lambda_c^+$ and $\bar{\Lambda}_c^-$, and likewise between $\Lambda_b^0$ and $\bar{\Lambda}_b^0$, offers a potential experimental probe of the magnetic field and its time evolution in relativistic heavy-ion collisions at RHIC and LHC energies.

Besides, the initial heavy-quark polarization can also be caused by other sources, such as spin correlations in heavy-quark production in hard scattering processes \cite{Zhang:2019xya,Li:2022pyw}, coupling of heavy-quark spin to the vortical QGP medium \cite{Ayala:2020ndx,Becattini:2007sr}, etc. Though the magnitude of the splitting between the charge-odd polarization of the heavy-flavor hyperons depends on the initial polarization of the heavy quark, the characteristic charge-odd structure of the precession signal and the heavy-flavor scaling $\Theta_c/\Theta_b=2m_b/m_c$ remains independent of its absolute magnitude. Consequently, the magnetic-field-induced rotation and its subsequent effects allow us to probe the longevity of the electromagnetic field in the thermalized deconfined QCD matter.

\section{Results and Discussion}

\begin{figure}[t]
\centering
\includegraphics[width=0.99\linewidth]{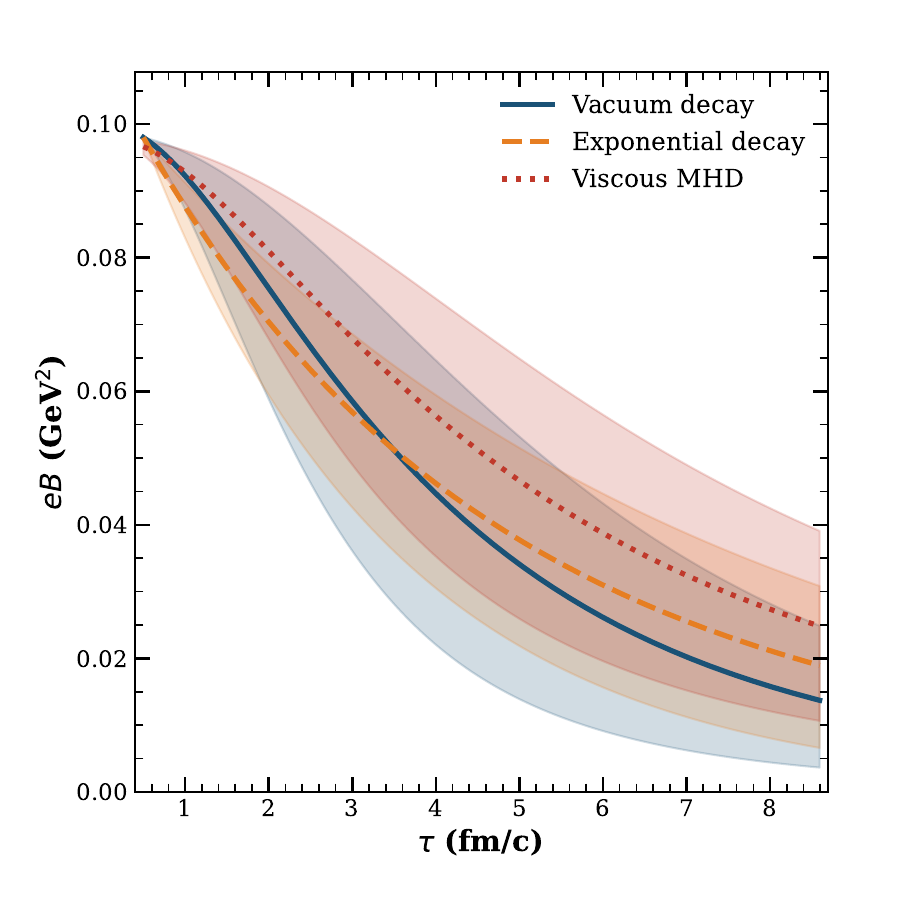}
\caption{Evolution of external magnetic fields for three different evolution scenarios as a function of proper time.}
\label{fig1}
\end{figure}

\begin{figure*}[t]
\centering
\includegraphics[width=0.45\linewidth]{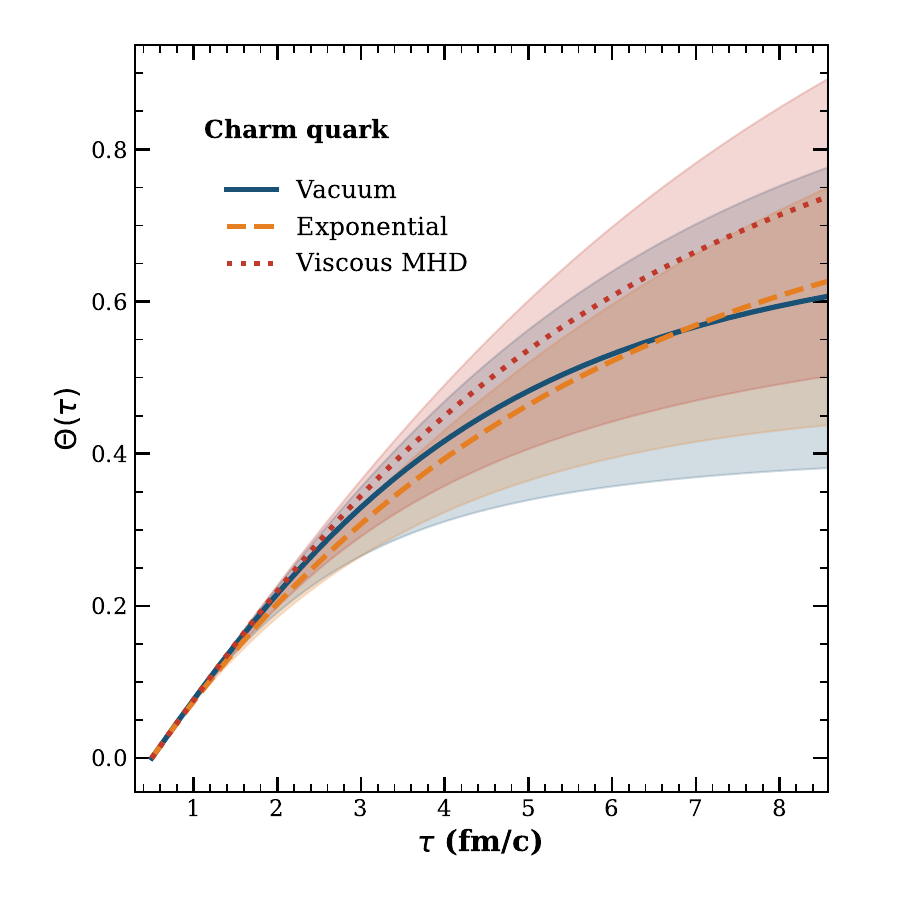}
\includegraphics[width=0.45\linewidth]{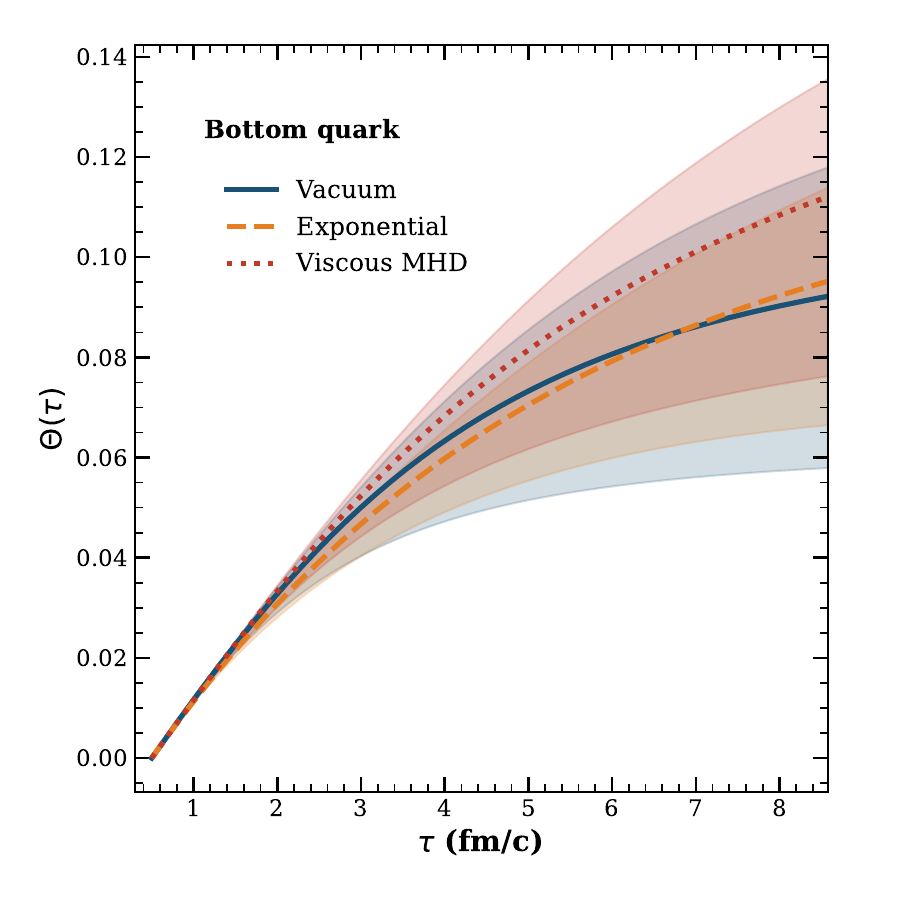}
\caption{Accumulated charm quark (left panel) and bottom quark (right panel) Larmor frequency as a function of proper time for different magnetic-field evolution scenarios.}
\label{fig2}
\end{figure*}

Figure~\ref{fig1} demonstrates the magnetic field evolution with time corresponding to the magnetic-field profiles discussed in Eq.~\ref{eq:Bprofiles}. Initially, a rapid decay of the magnetic field with time is observed, which tends to slow down as time progresses. The decay of the magnetic field in vacuum can be considered the baseline, as it lacks an external element that could cause any kind of changes in the magnetic field. Such as the exponential decay profile, which more likely treats the QGP as a perfect fluid, initially shows a faster decay of the magnetic field than the vacuum profile. In this profile, the initial rapid decay is the consequence of the magnetic energy invested in aligning the medium's constituents in a particular direction. Later, those aligned charged particles sustain the magnetic field, leading to a relatively longer lifetime compared to the decay rate in vacuum. The viscous MHD profile is parameterized to account for the initial magnetization of the medium. Therefore, the magnetic field decay rate in the medium is effectively reduced relative to vacuum and exponential decay profiles.\\

Now, we examine the accumulated Larmor phase acquired by a heavy quark during its propagation through the deconfined medium. We have chosen the initial magnetic field of $5m_{\pi}^2$ and varied $\tau_{B}$ in between 3--7 fm as per the LHC energy scenario, which enters into the results as uncertainty. The study is performed for all three magnetic field profiles. As shown in Fig.~\ref{fig2}, we obtain accumulated phases for the heavy quark of order $\Theta \sim 0.3-1.0$ for the charm quark, whereas, for the bottom quark, the accumulated phase is substantially small owing to its higher mass. For the charm quark, these values are large enough to induce a substantial rotation, leading to the polarization of a vector meson prior to or at hadronization. It implies that long-lived magnetic fields can modify the polarization pattern inherited by hadrons with at least one constituent that is a heavy quark or antiquark (charm or bottom). In addition, no substantial magnetic field dependence can be observed due to the uncertainty in the magnetic field decay timescale.\\

\begin{figure*}[t]
\centering
\includegraphics[width=0.45\linewidth]{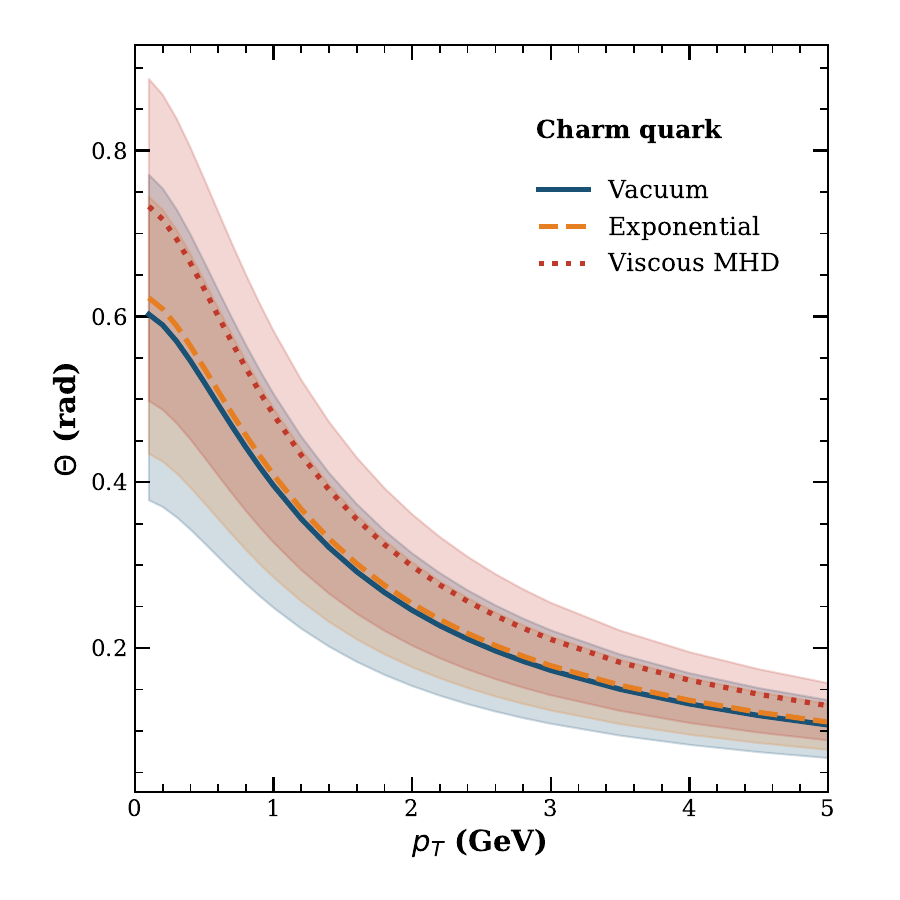}
\includegraphics[width=0.45\linewidth]{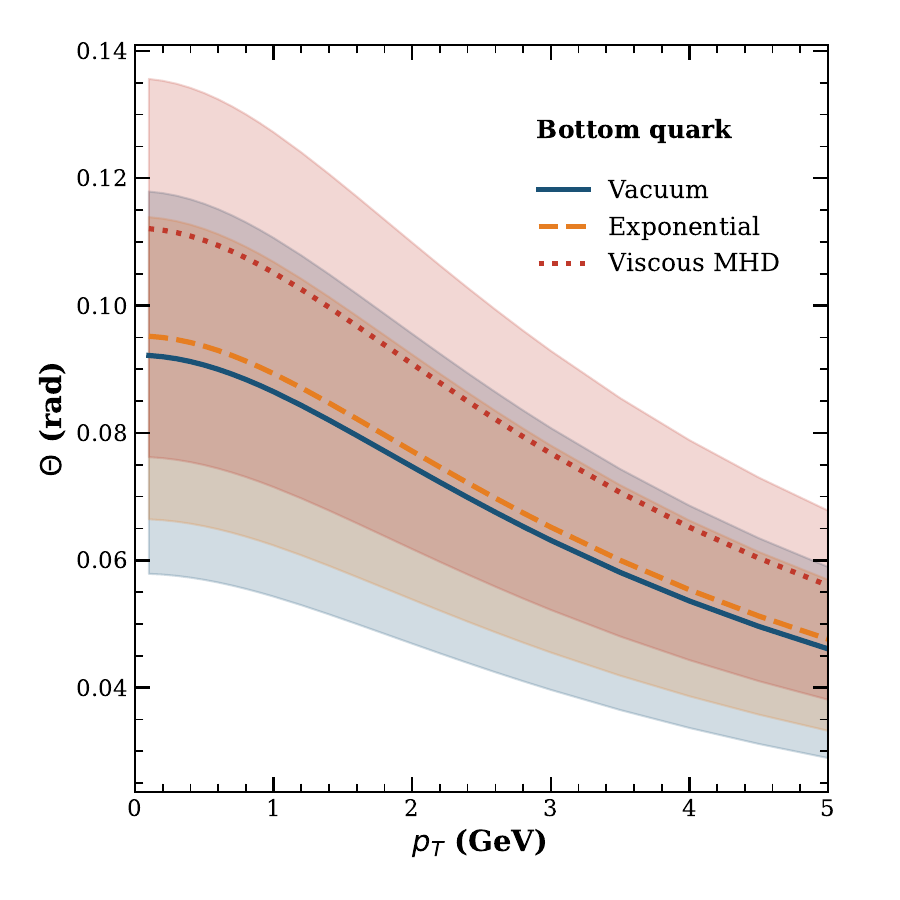}
\caption{Accumulated charm quark (left panel) and bottom quark (right panel) Larmor phase as a function of transverse momentum for different magnetic-field evolution scenarios.}
\label{fig3}
\end{figure*}

Figure~\ref{fig3} shows the accumulated Larmor phase as a function of transverse momentum for both charm and bottom quark in the left and right panel respectively, for all three decay profiles. The initial magnetic field is taken as $5m_{\pi}^2$.  As the Larmor frequency is inversely proportional to the net energy of the heavy quarks, in Fig.~\ref{fig3}, the transverse-momentum ($p_{T}$) dependence of the precession phase experiences weaker alignment at high $p_{T}$ than low $p_{T}$ quarks. As a result, the accumulated phase decreases monotonically with increasing transverse momentum. However, the decrease is substantial for the charm quark than the bottom quark.

\begin{figure*}[t]
\centering
\includegraphics[width=0.99\linewidth]{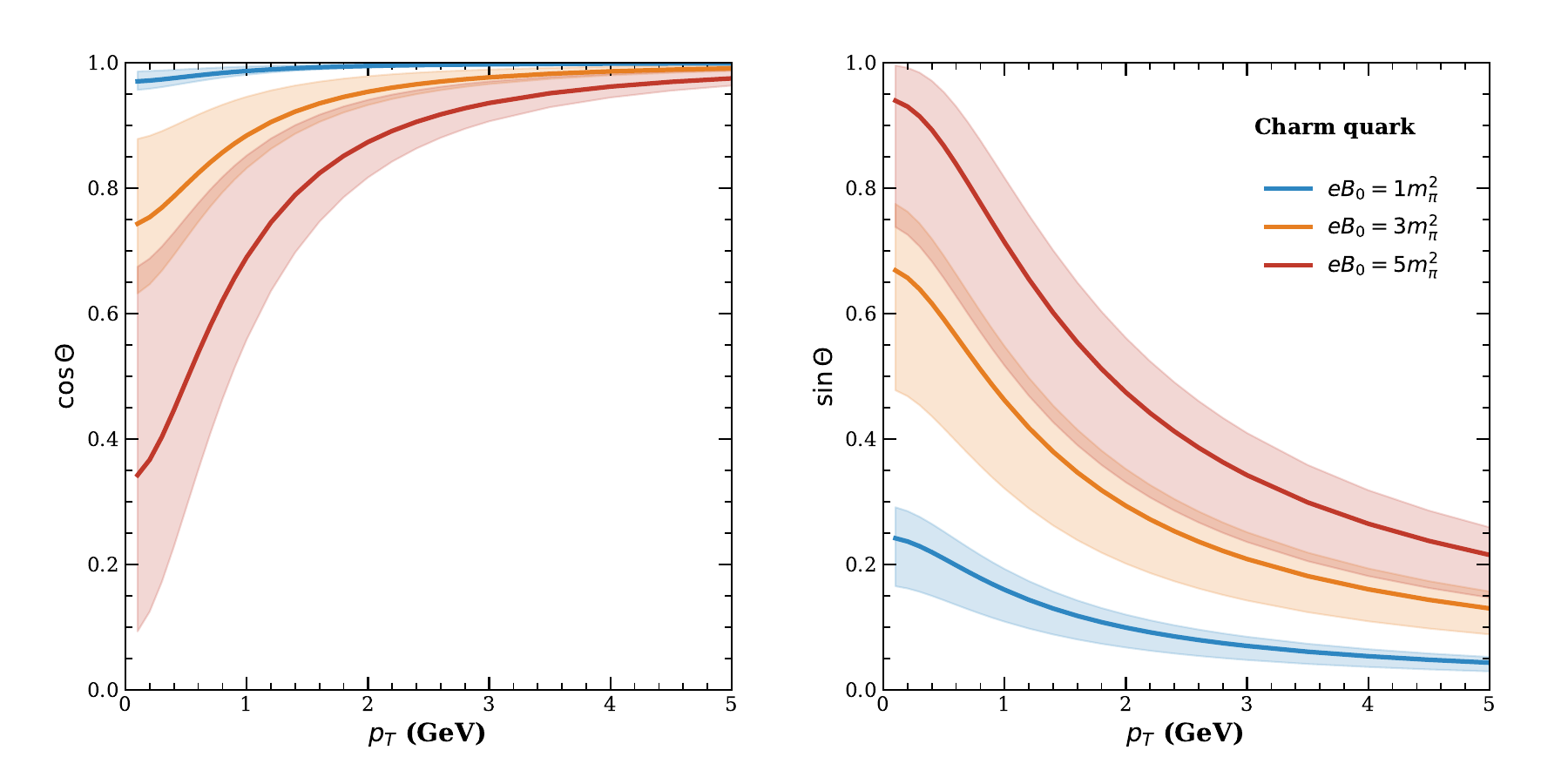}
\caption{Cosine (left panel)  and sine (right panel) of the accumulated charm quark Larmor phase as a function of transverse momentum
for different magnetic-field lifetimes in the viscous MHD evolution scenario.}
\label{fig4}
\end{figure*}

\begin{figure*}[t]
\centering
\includegraphics[width=0.99\linewidth]{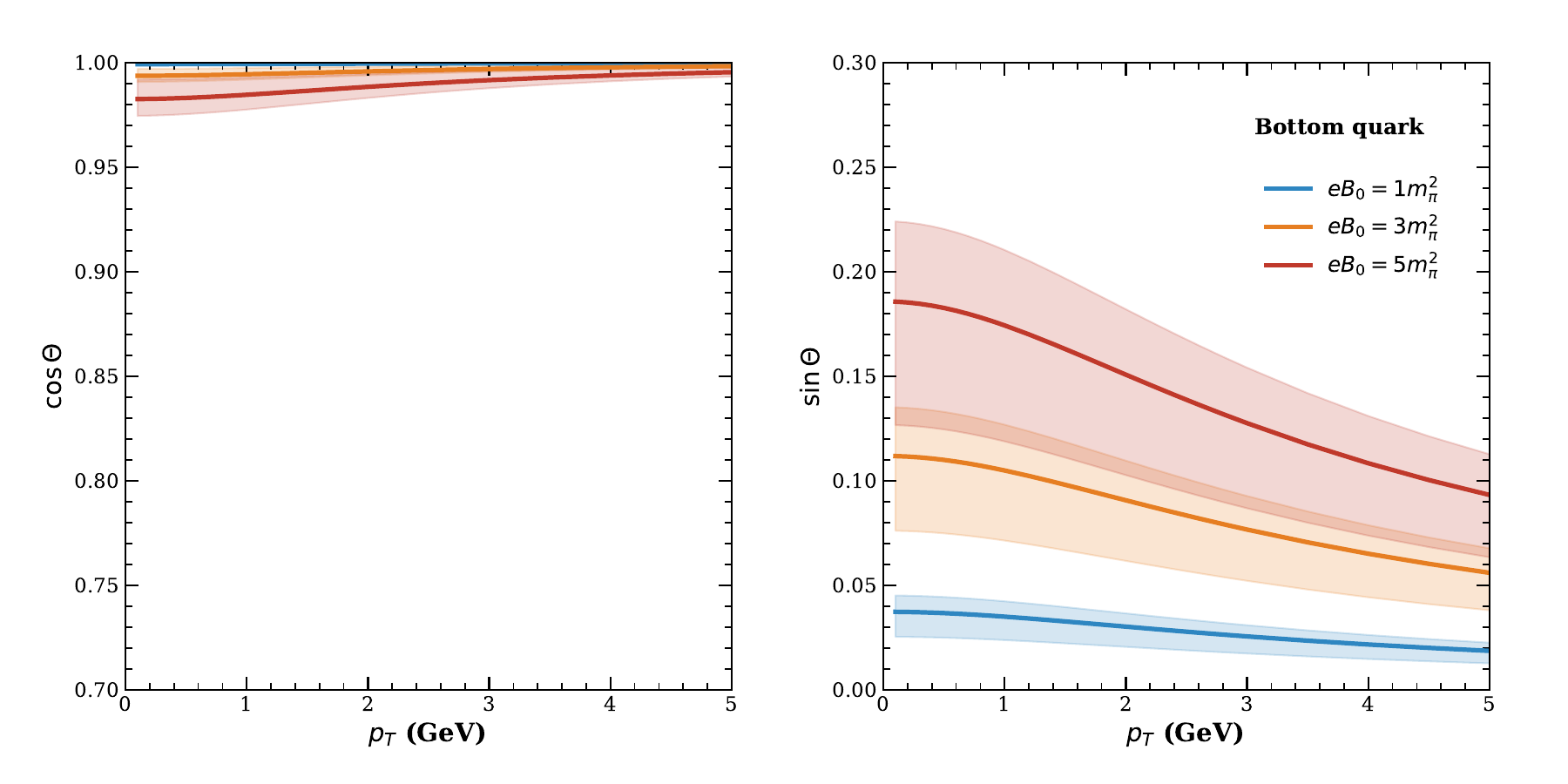}
\caption{Cosine (left panel)  and sine (right panel) of the accumulated bottom quark Larmor phase as a function of transverse momentum
for different magnetic-field lifetimes in the viscous MHD evolution scenario.}
\label{fig5}
\end{figure*}

Figure~\ref{fig4} shows the transverse momentum dependence of $cos\Theta$ and $sin\Theta$ for the charm quarks for three different initial magnetic field values, all decaying with the same viscous MHD decay profile. In this case, we see that at intermediate $p_{\rm T}$, both the $cos\Theta$ and $sin\Theta$ have significant contribution for a higher initial magnetic field case. Thus, for a heavy-ion collision at LHC, Larmor precession can shave substantial impact on the heavy quarks which can be reflected in the transverse momentum dependence of the longitudinal and transverse polarization observables. On the other hand, figure~\ref{fig5} shows the transverse momentum dependence of $cos\Theta$ and $sin\Theta$ for the bottom quarks for three different initial magnetic field values decaying with viscous MHD scenario. Here, the Larmor precession doesn't have as large of an impact as in the case of charm quark, but still can affect the final state longitudinal and transverse polarization to an extent.

\begin{figure*}[t]
\centering
\includegraphics[width=0.99\linewidth]{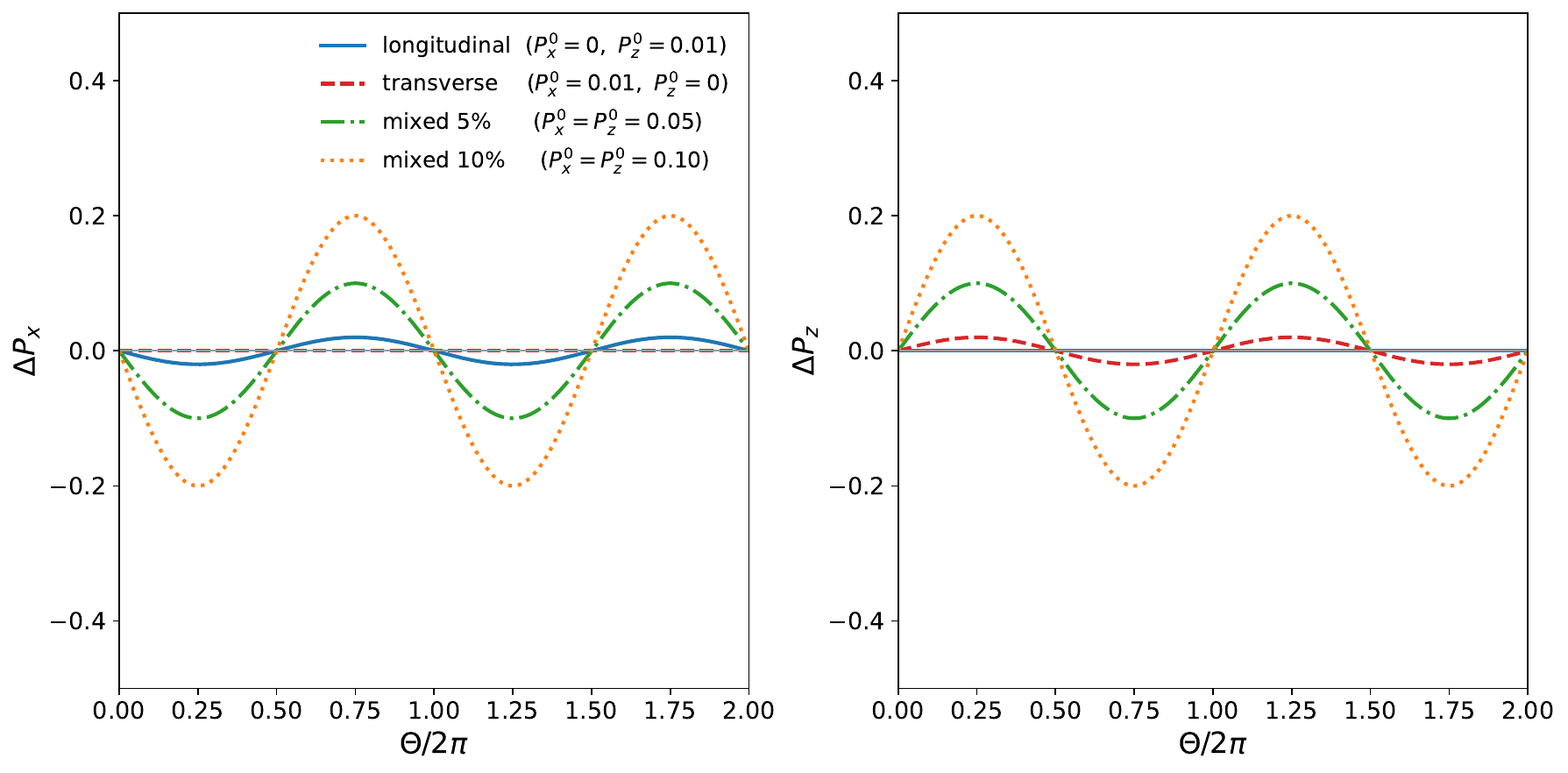}
\caption{Polarization splitting of heavy quarks as function of scaled accumulated phase for different initial values of $P_{x}^{0}$ and $P_{z}^{0}$.}
\label{fig6}
\end{figure*}

Finally, figure~\ref{fig6} shows the qualitative charge-odd polarization splitting between a heavy quark and its antiquark, $\Delta P_x$ and $\Delta P_z$, as a function
of the accumulated Larmor phase $\Theta/2\pi$, for representative initial polarizations. The quark and antiquark are assumed to start with the same polarization, as expected for vorticity-driven
production, and their opposite charges cause them to precess in opposite directions. Two features are physically important. First, the splitting is a pure
sine of the accumulated phase, it vanishes at $\Theta = 0$, peaks at $\Theta = \pi/2$, and reverses sign beyond it. Since the quark and antiquark evolve identically at zero phase, any charge-odd observable must vanish there, and because thermal vorticity is charge-even it cannot
generate such a splitting. Second, the $x$-component is controlled by the initial longitudinal polarization and the $z$-component by the initial transverse polarization, reflecting the fact that Larmor precession rotates the polarization vector in the $x$--$z$ plane. The pattern of which component shows a splitting thus encodes the orientation of the initial polarization, while the amplitude scales linearly with its magnitude.

An advantage of the charm or bottom sector arises from the intrinsic spin dynamics of heavy quarks. In a strongly interacting medium, repeated scattering processes can gradually randomize the spin orientation of light and strange quarks, reducing their sensitivity to the earliest stages of the collision. Heavy quarks, on the other hand, experience significantly suppressed spin-flip transitions due to their large mass, resulting in substantially longer spin-relaxation times. Consequently, charm quarks are expected to preserve information about their initial spin state over a much larger fraction of the QGP evolution. This property makes charm polarization particularly sensitive to coherent electromagnetic effects accumulated throughout the lifetime of the medium. In this sense, charm quarks act as effective ``spin memories'' of the early-time quark--gluon plasma, retaining information about the strongest magnetic fields generated immediately after the collision. The resulting polarization observables therefore probe the integrated electromagnetic history of the system, whereas light-flavor polarization is generally more susceptible to late-time medium effects, hadronic rescattering, and spin depolarization mechanisms. 

An important consequence of the present mechanism is the existence of a null test for long-lived magnetic fields in heavy-ion collisions. Larmor precession requires a finite accumulation time and therefore depends directly on the lifetime of the magnetic field. If the magnetic field decays rapidly after the collision, corresponding to $\tau_B \ll 1$ fm/$c$, the accumulated charm-quark precession phase remains negligible and no significant spin rotation can develop before hadronization. In this limit, charm and anti-charm quarks experience essentially identical spin evolution, leading to nearly identical polarization patterns for $\Lambda_c^+$ and $\bar{\Lambda}_c^-$ baryons. Conversely, if the magnetic field survives for several fm/$c$, the opposite electric charges of charm and anti-charm quarks generate opposite Larmor rotations, producing a characteristic charge-dependent polarization splitting between $\Lambda_c^+$ and $\bar{\Lambda}_c^-$. The observation of such a splitting would therefore constitute direct evidence for a non-vanishing time-integrated magnetic field, while its absence would imply strong constraints on the magnetic-field lifetime. In this sense, the polarization difference between charmed baryons and anti-baryons serves as an experimentally accessible order parameter for long-lived electromagnetic fields in the quark--gluon plasma.

\section{Conclusion and Outlook}
The central outcome of this work is the identification of charm and bottom quark Larmor precession as a clean and direct probe of long-lived magnetic fields in relativistic heavy-ion collisions. Unlike the strange sector, where medium-dependent masses, feed-down contributions, and competing polarization mechanisms complicate the interpretation of electromagnetic effects, the large and approximately constant heavy quark mass allows the accumulated precession phase to be directly related to the time-integrated magnetic field experienced during the QGP evolution. We find that realistic magnetic fields can generate sizeable spin rotations prior to hadronization, leading to characteristic charge-dependent polarization patterns in $\Lambda_c^+$ and $\bar{\Lambda}_c^-$ baryons and a distinctive transverse-momentum dependence arising from relativistic suppression. These signatures provide a falsifiable experimental test of long-lived magnetic fields and open a new avenue for studying the electromagnetic properties of the quark--gluon plasma through heavy-flavor spin observables. Future measurements of charmed-baryon polarization at ALICE and LHCb will therefore offer a unique opportunity to directly constrain the magnetic-field lifetime in heavy-ion collisions. Since here, the Larmor precession acts at the partonic level before hadronization, its influence may extend beyond charmed baryons to other heavy-flavor spin observables. Whether analogous signatures can appear in vector-meson spin-alignment measurements requires a dedicated treatment of the spin density matrix and hadronization process, and remains an interesting direction for future investigation.\\

An important aspect for experimental observation is the event-by-event fluctuation of the magnetic field. While the magnitude and orientation of the field vary from event to event due to fluctuations in the nuclear overlap geometry and initial state, the charge-odd nature of the predicted polarization splitting ensures a robust signal. Specifically, the polarization difference \(\Delta P = P_{Q} - P_{\bar{Q}}\) is odd under charge conjugation and therefore survives event averaging, unlike charge-even contributions from vorticity. Thus, while the absolute polarization magnitude may be reduced by fluctuations, the charge-dependent splitting remains a clean experimental observable. A quantitative assessment of fluctuation effects lies beyond the scope of this work but constitutes an important direction for future investigation.\\

The present results suggest a new strategy for probing magnetic fields in heavy-ion collisions. Rather than attempting to infer electromagnetic effects indirectly through modifications of light-hadron observables, one can exploit the spin dynamics of heavy quarks as a direct electromagnetic probe. Because the accumulated phase is controlled primarily by the integrated magnetic field and only weakly affected by medium-dependent mass corrections, the charm and bottom sectors offer a substantially cleaner environment for isolating electromagnetic effects than the strange sector.\\

\section*{Acknowledgement}
DS acknowledges the support from the postdoctoral fellowship of the DGAPA UNAM, México. Captain R. Singh acknowledges the Department of Atomic Energy (DAE), India, for financial support.
\\

\end{document}